\documentstyle[prl,epsfig,aps,multicol]{revtex}
\title
{\bf Breathers in a Discrete Nonlinear Schr\"{o}dinger type Model : Exact 
Stability Results.}
\author{\bf{Avijit Lahiri}$^{1}$, {Subhendu Panda}$^{2}$ and {Tarun K.
Roy}$^{2}$}
\address{\bf$^{1}$Dept of Physics, Vidyasagar Evening College, Kolkata 700 006, INDIA \\
\bf$^{2}$Saha Institute of Nuclear Physics, 1/AF, Bidhannagar, Kolkata 700 064, INDIA}
\begin{document}
\maketitle
\begin{abstract}
Exact stability analysis of 1-site breathers in an NLS type model (ref. [4],
see below) indicates destabilisation through a growth rate becoming positive
as a stability border is crossed, while above a critical spatial decay rate 
($\lambda_c$) the breather is always unstable. Similar analysis on 2-site 
breathers with phase difference $\pi$ (see below) indicates destabilisation
through {\it Krein collision} of a pair of growth rates. All eigenmodes can be 
exactly calculated and conform to numerical estimates. 
\\ \\  
PACS number(s): 05.45.-a, 63.20.Pw, 63.20.Ry
\end{abstract}
\begin{multicols}{2}
\noindent The question of dynamical stability of discrete breathers has been addressed in
the literature and a number of useful results are now known. The stability of
a monochromatic breather depends on the eigenvalues of a symplectic matrix, the
so called Floquet matrix. Aubry \cite{aubry} introduced a novel extension to the concept
of the Floquet matrix , thereby greatly simplifying the task of computing the
Floquet multipliers and of looking for the onset of instability. In particular, he established that, one-site breathers sufficiently close to the 
`anti-integrable' limit are dynamically stable. However, few exact results
relating to the multipliers and the associated eigenmodes are known (\cite{fw},
\cite{krb};
see references therein for a background to the problem) notably 
because the breather solutions around which the linearisation is to be performed are 
themselves obtained numerically, even though with high precision. In this 
context it is useful to have models that admit of {\it exact} breather solutions as
also exact stability results. In \cite{lpr} we introduced a number of piecewise smooth
(PWS) models including a nonlinear Schr\"{o}dinger type one and worked out
exact one-site breather solutions where one of the sites is excited to a value 
above a {\it threshold} while all other sites remain below the threshold ( we denote
these as respectively {\it high} and {\it low}). In the present paper we work out exact
stability results for these breathers as also for {\it 2-site} breathers where two
sites with an arbitrary intervening gap are {\it high} while all other sites 
are {\it low}. While
the destabilisation of 1-site breathers always occur through the growth rate of
a mode becoming positive, the 2-site breathers get distabilised through a 
{\it Krein collision} of the growth rates (see below). \\
\noindent More precisely, the PWS Schr\"{o}dinger type model introduced in \cite{lpr} is
\begin{eqnarray}
i{{d\psi_n}\over {dt}} + V (\psi_{n+1} + \psi_{n-1}) = \gamma \psi_n f(\psi_n),
(n=0,\pm 1,...)   \label{eq:one} \\
f(z) = -(1-a/|z|) \Theta(|z|-a),~~~~~~~~~~~~~~~~~~~~~~~~~~~~~ \label{eq:two}
\end{eqnarray}
\noindent where $a(>0),~V (>0),~ \gamma$ are appropriate constants and $\Theta$
denotes the Heaviside step function. We scale the nearest neighbour coupling 
strength and the threshold parameter to $V=1$, $a=1$. Fig.1 compares $f(z)$ of (\ref{eq:two}) with
the more usual form $ f(z) = -|z|^2 $ of the discrete non-linear 
Schr\"{o}dinger equation. The one-site breather solution centred at $n=0$ is
then

\begin{eqnarray}
\bar{\psi}_n&=&\bar{\phi}_ne^{-i\omega t},  \label{eq:three} \\
\bar{\phi}_n&=&\frac{\gamma\lambda^{|n|}}{\gamma+\lambda-1/\lambda} \label{eq:four}.
\end{eqnarray}

\begin{figure}
\begin{center}
\epsfig{figure=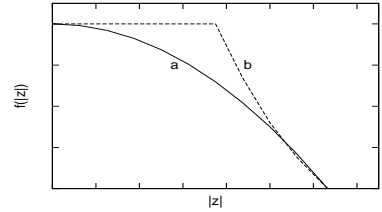,height=3cm,width=5cm}
\end{center}
\caption{Comparison of function $f(z)$ (a) of the usual form $f(z)=
-|z|^2$ with (b) of Eq.(2).} 
\end{figure}

The spatial decay rate $\lambda~(|\lambda|<1)$ and the breather frequency 
$\omega$ satisfy 
\begin{equation}
|\omega|>2,~ \omega=-(\lambda+1/\lambda),~ \omega\gamma<0,~ 
|\gamma|>1+\frac{1}{|\lambda|},  \label{eq:five}
\end{equation}
(note: Equations (10), (11) in \cite{lpr} expressing the above relations
contained errors due to oversight, which we regret).We distinguish
between breathers of {\it type A} $(\lambda>0, \omega<-2)$ and {\it type B}
$(\lambda<0, \omega>2)$), referring to the former in presenting
our results below (results for type B are obtained analogously). The overbar in 
(\ref{eq:three}),(\ref{eq:four}) is used to distinguish the breather
solution from neighbouring perturbed solutions (see below). \\
We consider these perturbations in the `rotating frame' in the form 
\begin{equation}
\psi_n = (\bar{\phi}_n + u_n(t))e^{-i\omega t}. \label{eq:six} 
\end{equation}
Spliting $u_n$ into real and imaginary parts, $u_n=x_n+iy_n$, one observes that
the time variations of $x_n,~y_n$ are obtained from the Hamiltonian 
\begin{eqnarray}
H = \sum_n {\{(x_{n+1}x_n + y_{n+1}y_n) + {{\omega}\over{2}} ({x_n}^2+{y_n}^2)\}} + \nonumber \\
  ~~~~~~~~~~~~~~~~~+{{\gamma}\over{2}}\{{x_0}^2+{y_0}^2(1-{{1}\over{b}})\} \nonumber, 
\end{eqnarray}
where $b=\frac{\gamma}{\gamma+\lambda-\frac{1}{\lambda}}$. Eigenvalues of the
matrix $L$ determining the evolution of the $x_n$'s and $y_n$'s  will be termed 
{\it growth rates} (we denote a typical growth rate by the symbol $p$) while a more 
convenient approach in the present context is to eleminate the $y_n$'s or 
$x_n$'s, arriving at
\begin{equation}
\ddot{X}=AX,~~~~\ddot{Y}=A^TY~~~~~~~(\mu \neq 0), \label{eq:eight}
\end{equation}
where we define $X=(........x_{-2},x_{-1},x_0,x_1,x_2,.......)^T $, 
$Y=(.........y_{-2},y_{-1},y_0,y_1,y_2,........)^T $, and $A$ is a banded matrix
with elements $a_{mn}~(-\infty<m,n<\infty)$ differing from zero only when 
$|m-n| \leq 2 $ :  \\

$~~~~~~~~a_{0,0} = 2(\gamma \lambda -2 -\lambda^2),$ \\
$~~~~~~~~~~~~a_{0,1} = a_{0,-1} = 3\lambda +1/\lambda,$ \\
$~~~~~~~~~~~~a_{1,0} = a_{-1,0} = 2\lambda +2/\lambda -\gamma,$ \\
$~~~~~~~~~~~~a_{n,n} = -(\lambda^2 +1/\lambda^2 +4),~~~ (n\ne 0),$ \\
$~~~~~~~~~~~~a_{n,n+1} = a_{n+1,n} = 2(\lambda +1/\lambda),~~~ (n \neq 0,-1),$ \\
$~~~~~~~~~~~~a_{n,n+2} = a_{n+2,n}=-1 $.\\ \\
\noindent The eigenvalues $(\mu)$ of $A$ are related to growth rates $(p)$ as $\mu = p^2$.
The spectrum of eigenvalues $\mu$ and the corresponding eigenmodes can all be
worked out exactly for any given parameter values $ \lambda,~\gamma$ satisfying
(\ref{eq:five}). There exists a band of eigenvalues $-(\omega-2)^2<\mu<
-(\omega+2)^2 $ with {\it extended} eigenmodes, there being a {\it symmetric} 
and an 
{\it antisymmetric} eigenmode associated with each eigenvalue. 
Additionally, there exists a single {\it localised} symmetric eigenmode for 
$\gamma>\gamma_1(\lambda)$ where
\begin{eqnarray}
{\gamma_1}(\lambda)=\frac{(\lambda-\frac{1}{\lambda})\sqrt{((\lambda
+\frac{1}{\lambda})^2-2(\lambda+\frac{1}{\lambda}))}}{(\lambda-
\frac{1}{\lambda})+\sqrt{((\lambda+\frac{1}{\lambda})^2-
2(\lambda+\frac{1}{\lambda})}}. \nonumber
\end{eqnarray} 
\noindent For any given $\lambda,~\gamma_1$ is thus the {\it threshold} for the localised mode:
for $\gamma<\gamma_1$ there is no localised mode and the band edges are empty,
while for $\gamma=\gamma_1$ an eigenmode appears at the band-edge, and 
dissociates from the band as the localised mode for $\gamma>\gamma_1$. One
notes in this context that there always exists a trivial eigenmode $\mu=0$
with $x_n=0,~y_n\propto \bar{\phi}_n$, which is to be distinguished 
from the eigenvectors of $A,~A^T$ with $\mu=0$, the latter 
corresponding to a spurious mode in the present
context. While a typical localised eigenvector of $A$ is of the form
\begin{eqnarray}
x_n=\alpha \chi_1^{|n|} + \beta \chi_2^{|n|}, \nonumber
\end{eqnarray}
where the spatial decay rates $\chi_1,~\chi_2$ (each of modulus $<1$) as also
the ratio $\alpha/\beta $ can be obtained exactly, the eigenvector 
corresponding to $\mu=0$ looks like 
\begin{eqnarray}
x_n=(\alpha+\beta|n|)\lambda^{|n|}. \nonumber
\end{eqnarray}
The growth
rates for the localised mode are imaginary $(p=\pm i\sqrt|\mu|)$ so long as
$\gamma_1(\lambda)<\gamma<\bar{\gamma}(\lambda)$ where the transition value
$\bar{\gamma}$ can also be obtained exactly in the model: 
\begin{eqnarray}
\bar {\gamma} (\lambda) =  \frac{(1+4\lambda^2+\lambda^4)(1-\lambda^2)}{2\lambda^3}. \nonumber 
\end{eqnarray} 
\noindent As $\gamma$ crosses $\bar{\gamma}$ from below, the isolated eigenvalue $\mu$ crosses $\mu=0$ and 
becomes positive, giving rise to a positive growth rate $(p_+=\sqrt{\mu};$ this 
is associated with a negetive growth rate $p_{-}=-\sqrt{\mu}$, making up a pair)
and implying destabilisation of the breather.  
\begin{figure}
\begin{center}
\epsfig{figure=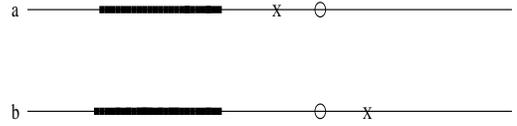,height=2cm,width=7cm}
\end{center}
\caption{Spectrum of eigenvalues of matrix $A$ for $(a)$ $\gamma_1(\lambda)<
\gamma<\bar{\gamma}(\lambda)$ and $(b)$ $\gamma>\bar{\gamma}(\lambda)$; cross
denotes eigenvalue for the isolated localised mode while thick black line denotes 
the band; circle corresponds to $\mu=0$.}
\end{figure}
\noindent Fig.2 depicts the disposition
of eigenvalues on either side of the destabilisation, while
\begin{figure}
\begin{center}
\epsfig{figure=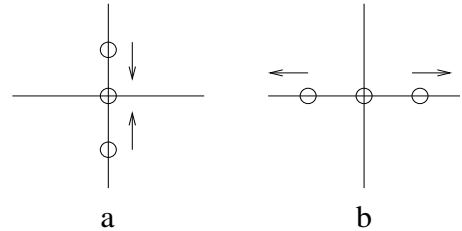,height=3cm,width=6cm}
\end{center}
\caption{Disposition of growth rates $p$ in the complex plane with $(a)$ 
$\gamma<\bar{\gamma}$ (b) $\gamma>\bar{\gamma}$}
\end{figure}
\noindent Fig.3 indicates how the disposition of growth rates in the complex plane 
changes across the destabilisation border.
\begin{figure}
\begin{center}
\epsfig{figure=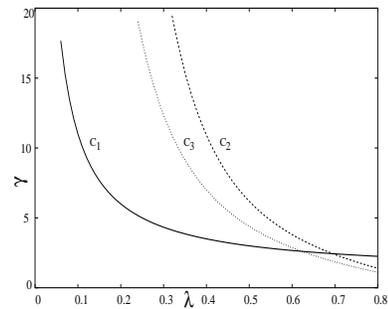,height=4cm,width=5cm}
\end{center}
\caption{Curves $C_1$ $(\gamma=\gamma_0(\lambda))$, $C_2~ (\gamma=\bar{\gamma}
(\lambda))$ and $C_3~(\gamma=\gamma_1(\lambda))$ in the $\lambda-\gamma$ plane;
the point of intersection of $C_1$ and $C_2$ corresponds to $\lambda=\lambda_c
\approx 0.6948$}
\end{figure}
\noindent Fig.4 depicts the $\lambda-\gamma$ parameter space for the model 
together with the
curves $C_1~(\gamma=\gamma_0(\lambda)~\equiv~ 1+\frac{1}{\lambda}),~ 
C_2~(\gamma=\bar{\gamma}(\lambda)),~ C_3~(\gamma=\gamma_1(\lambda))$. The breather exists
only for points lying above $C_1$ and is stable only for points lying below 
$C_2$, $i.e.,$ for $(\lambda,~\gamma)$ lying in between $C_1$ and $C_2$ 
the model 
admits of a stable breather solution. The curve $C_3$ gives the {\it threshold} for
the localised mode. Interestingly, the model predicts the existence of a critical
value $(\lambda_c)$ of the spatial decay rate corresponding to the intersection
of $C_1$ and $C_2$ $(\lambda_c\approx 0.6948)$ so that the breathers with 
$\lambda>\lambda_c$ are unstable for all values of the strength of 
non-linearity $\gamma~(>\gamma_0(\lambda))$. All these features resemble those
of `internal' or `breathing' modes of DNLS breathers whose relevance has
been emphasized in the literature (\cite{lks},\cite{jagcr}). However, the antisymmetric
`pinning' mode \cite{jagcr}, known to exist for DNLS breathers, is
marginal in the present PWS model $(\lambda=1,~\mu=0)$ and possesses a linear 
spatial growth. 
\begin{figure}
\begin{figure}
\epsfig{figure=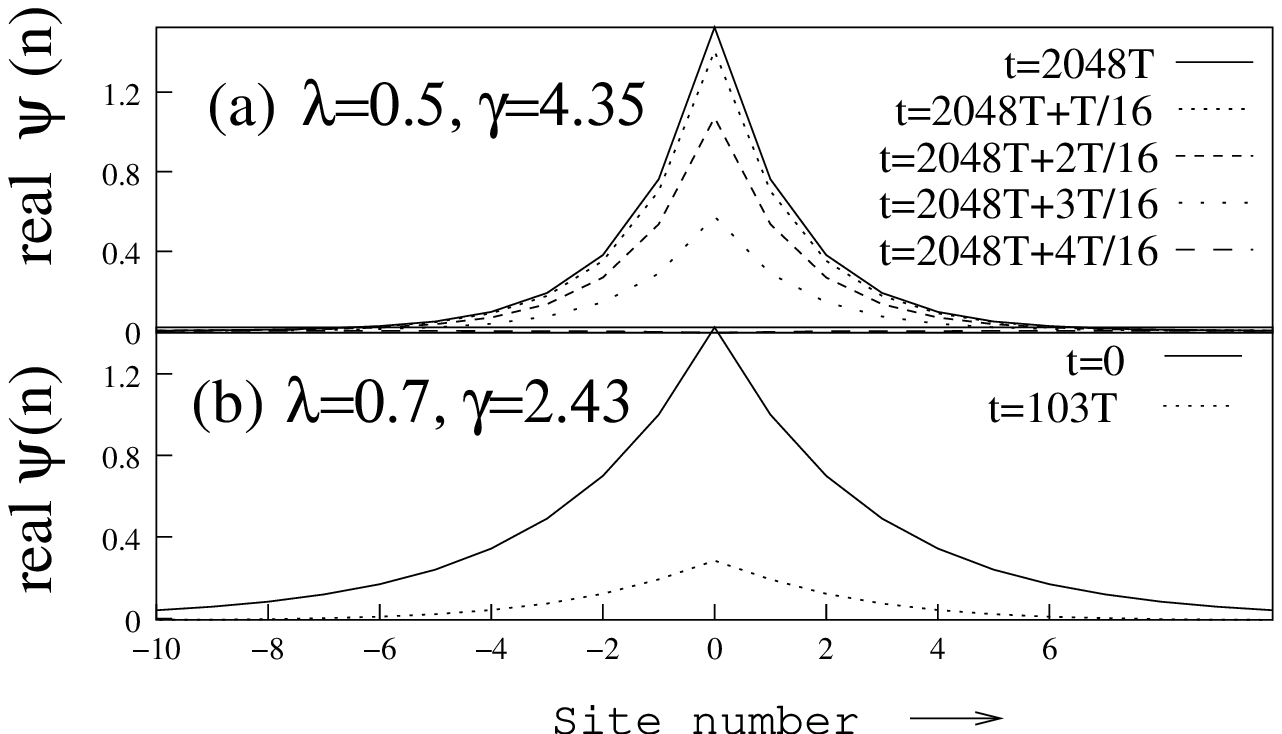,height=3cm,width=6cm}
\end{figure}
\caption{Numerical integration following (\ref{eq:one}), (\ref{eq:two}) of an initial profile given 
by (\ref{eq:four}) at $t=0$; (a) $\lambda=0.5,~\gamma=4.35,$ in between $C_1$ and $C_3$ of 
Fig.3 - profiles at values of $t$ spanning a quarter of a period $(T/4)$ 
starting from $2048T$; (b) $\lambda=0.7$ (just beyond intersection of $C_1$ and $C_2$ in Fig.3), and $\gamma=2.43
(>\gamma_0 (\lambda))$.}
\end{figure}

\noindent We show in Fig.$5a$ the temporal evolution according to Eqs. (\ref{eq:one}),
(\ref{eq:two}) of a profile initially (t=0) coinciding with (\ref{eq:three}),
with $(\lambda,\gamma)$ in the region between $C_1$ and $C_2$ while Fig.5b shows 
the similar evolution for a breather with $\lambda>\lambda_c$, above $C_1$. One notes that the breather in Fig. 5a performs stable oscillations while that in
Fig. 5b breaks up within a short time, confirming our results. In this context
see also Figs. 4,5 of \cite{lpr}.  \\
The eigenmodes can be obtained exactly in the model.
\noindent For instance, Fig.6
shows the localised eigenmode for $\lambda=0.5,~\gamma=6$ calculated from
the theory as compared with the eigenmode obtained numerically by
diagonalising the matrix $A$. 
\begin{figure}
\begin{center}
\epsfig{figure=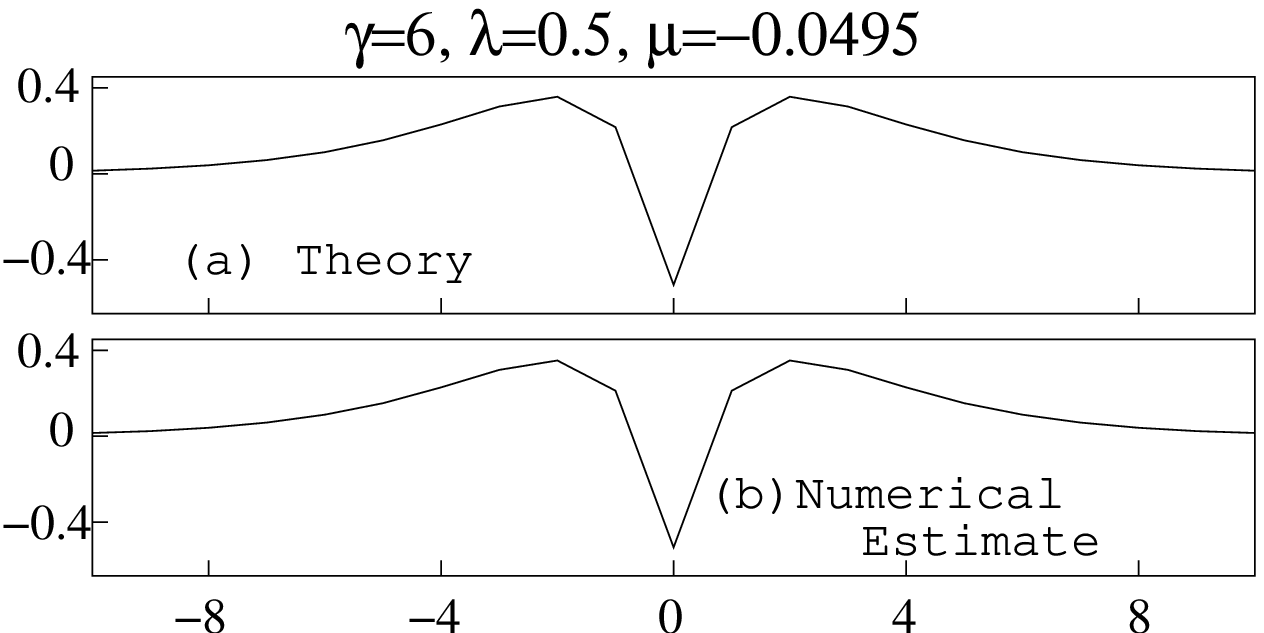,height=3cm,width=6cm}
\end{center}
\caption{Eigenmodes obtained (a) from exact analysis and (b) by diagonalising the matrix A}
\end{figure}
\noindent As we mentioned in \cite{lpr}, the PWS model (\ref{eq:one}), (\ref{eq:two}) is a 
veritable little `laboratory' yielding 
exact breather solutions of a wide variety. For instance, one can construct 
{\it two-site} breathers with various possible lengths of intervening gap $N$. One other
feature of crucial relevance is the relative {\it phase} between the two sites.
\noindent  This relates to the fact that (\ref{eq:four}) admits of an arbitrary phase factor
 $ e^{i\delta}$ that becomes relevant in 2-site or multisite breather 
solutions. 
Thus, considering a breather with two sites {\it high} ($i.e.,$ with $
|\psi_n|>1)$ and all the other sites {\it low} $(|\psi_n|<1)$ with the {\it high} sites having
opposite phases ($\delta=\pi$) one obtains a monochromatic breather solution
of the form (\ref{eq:three}) with the breather profile $\bar{\phi}_n$ now given by 
\begin{eqnarray} 
\bar{\phi}_n=\frac{b}{1-\lambda^N}(\lambda^n-\lambda^{N-n})~~~~~~ (0\leq n \leq N) \nonumber \\
            = b\lambda^{-n}~~~~~~~~~~~~~~~~~~~~~~~~~~~~~~~~~(n<0)  \nonumber \\
            = -b \lambda^{n-N}~~~~~~~~~~~~~~~~~~~~~~~~~~(n>N),  \label{eq:eleven}
\end{eqnarray}
where
$b=\frac{\gamma(1-\lambda^N)}{\gamma(1-\lambda^N)-(\lambda^{-1}-\lambda)}$
and $\lambda,~ \gamma$ are to satisfy $\gamma>\frac{1+\lambda^{-1}}{1-\lambda^N}.$
Here the two {\it high} sites have been chosen to be at $n=0$ and $n=N$ for any 
given separation $N$. 
It is now easy to construct the stability matrix $A$ of Eq.(\ref{eq:eight}) and to
obtain the spectrum of eigenvalues together with the exact eigenmodes. There
is once again a band extending from $-(\omega-2)^2$ to $-(\omega+2)^2$ (recall
that we are  considering only type {\it A} breathers, with $\omega<-2$) containing
extended eigenmodes, together with a number of isolated eigenvalues 
corresponding to localised modes. The latter appear above a certain
{\it threshold} value of $\gamma$ (for any given $\lambda$), and are now typically
{\it four} in number of which one is the trivial eigenvalue $\mu=0$ while the others 
lie in the range $-(\omega+2)^2<\mu<0 $.
Ordering these as $\mu_1<\mu_2<\mu_3<\mu_4=0,$ the eigenmodes corresponding to
$\mu_1$ and $\mu_4$ are {\it antisymmetric} (with respect to the sites $n=0,N$)
while the other two are {\it symmetric}. 
Fig.7 shows the four eigenmodes for $N=8,\gamma=4.9443,\lambda=0.5$.
For a given $N,~\lambda$, as $\gamma$ is made to increase from the threshold
value, a boundary $\gamma_H(\lambda)$ is reached when $\mu_2$ and $\mu_3$       become coincident
and then become complex.
In terms of the growth rates $p$ $(p=\pm\sqrt\mu)$
there occurs a 
Krein collision of a quartet of growth rates on the imaginary axis and the
subsequent spliting of the growth rates out of the complex plane, as depicted 
schematically in Fig.8. 
\begin{figure}
\vskip -.8cm
\begin{center}
\vskip 0.6cm
\epsfig{figure=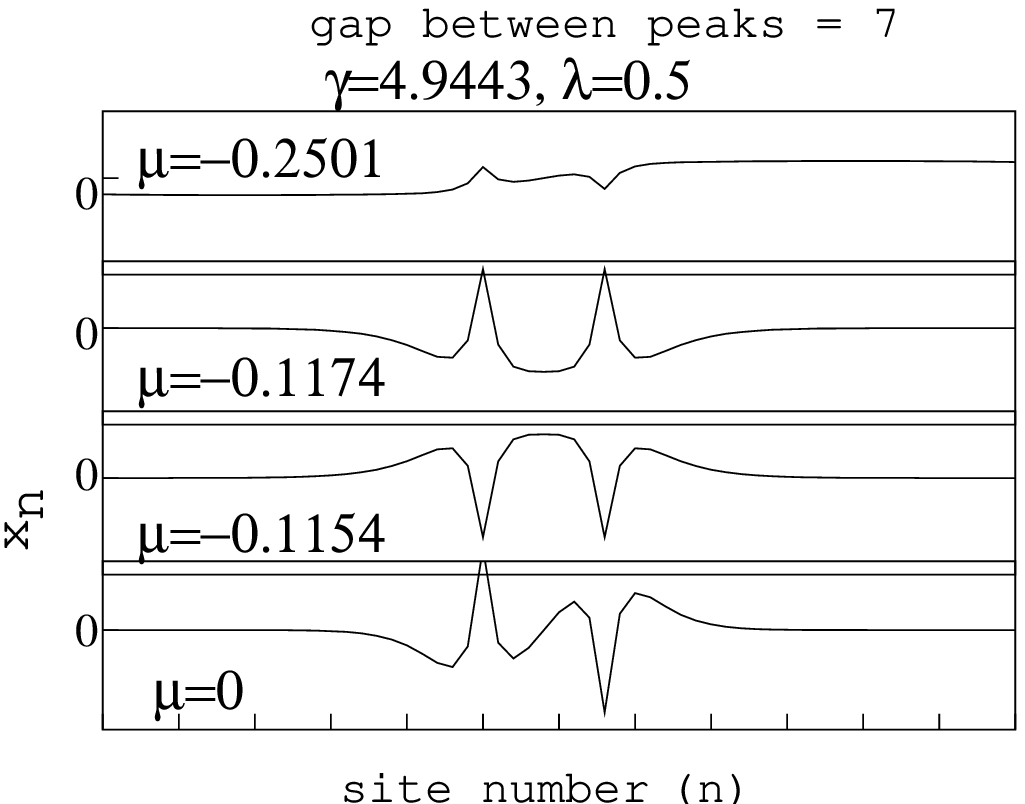,height=4cm,width=6cm}
\end{center}
\caption{Symmetric and antisymmetric localised eigenmodes for a 2-site breather.}
\end{figure}
\noindent In a model with sufficiently smooth $f(z)$ (ref.
Eq.\ref{eq:two}) this leads to the {\it Hamiltonian Hopf bifurcation} (\cite{ls-r}, see references
therein; the Krein collision has been discussed in \cite{aubry} in terms of the 
Floquet multipliers related to the breather solution - the latter are related
to our eigenvalues ($\mu$) and the corresponding growth rates $(p)$ in a simple
manner; there exists a large body of literature relating to 
Krein collision in symplectic mappings - see, $e.g.$, \cite{bf}). In the 
present context, the collision signifies a loss of stability. 
For instance, with $ N=8,
\lambda=0.5,$ the model predicts a loss of stability of the two-site breather 
through a Krein collision at $\gamma_H \approx 4.9444$.

\noindent Fig.9 depicts the time evolution of an initial profile chosen in
accordance with (\ref{eq:eleven}) 
with $\gamma$ on either side of the collision: one notes
that the breather indeed remains stable for $\gamma<\gamma_H$ and breaks up quickly for 
$\gamma>\gamma_H$. 

\begin{figure}
\begin{center}
\epsfig{figure=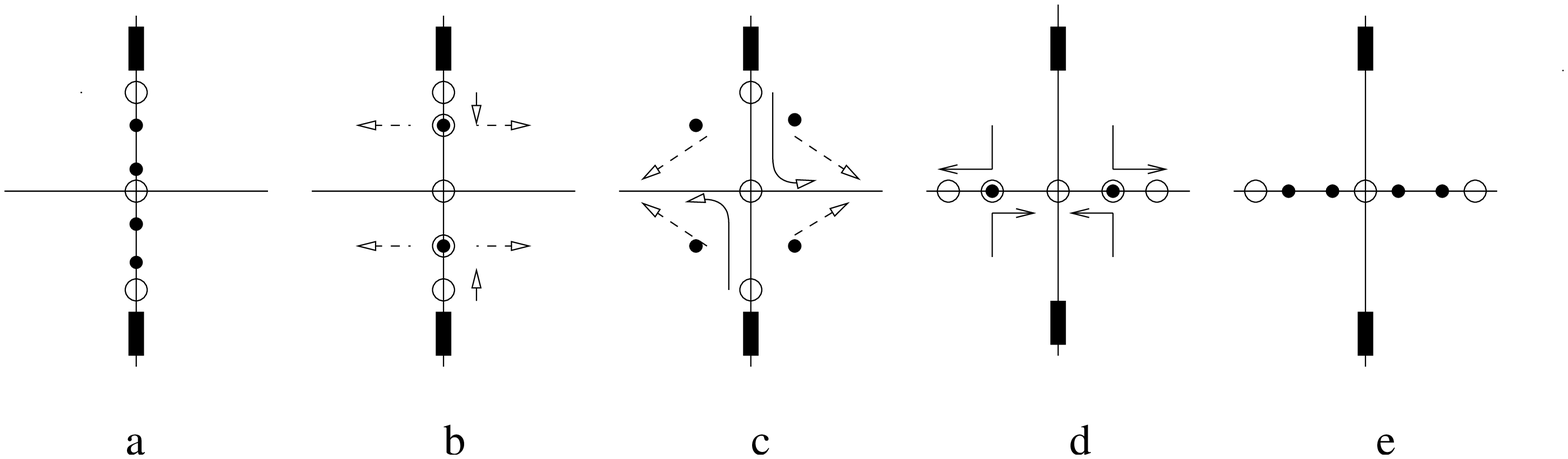,height=2cm,width=8cm}
\end{center}
\caption{Sequence of changes in the disposition of growth rates in the complex 
plane; thick black lines denote the band while the dots and circles 
correspond to localised modes; (a) before Krein collision($\gamma<\gamma_H$;
$\lambda$ fixed), (b) Krein collision $\gamma=\gamma_H$, (c) a quartet of complex 
growth rates resulting from the Krein collision, and a pair of imaginary 
growth rates approaching collision at $p=0$, (d) Krein collision `in the 
reverse', (e) all four growth rates of localised modes real.}
\end{figure}

\noindent As mentioned above, in Fig.8
we depict the sequence of changes in the disposition of the growth rates $p$
as $\gamma$ is made to increase through $\gamma_H$ for fixed $\lambda$.
One observes that at some value $\tilde{\gamma}(>\gamma_H)$ of $\gamma$, the 
root $\mu$ becomes positive (this
is similar to the crossing of a negative eigenvalue through the value zero at 
$\gamma = \bar{\gamma}(\lambda))$
with $\mu_2,~\mu_3$ remaining complex; subsequently at some value $\gamma=\gamma'_H$
the latter collide at a point on the positive real axis and move off in
opposite directions - Krein collision `in the reverse'. In the present context
of stability of the breathers, however, this reverse collision does not
have a direct relevance since the breather, having become unstable at $ \gamma=
\gamma_H$     
continues to remain unstable even at $\gamma=\tilde{\gamma}$ and $\gamma=\gamma'_H$.
For large $N$, one finds that $\gamma_H,~\tilde{\gamma}$ and $\gamma'_H$ are all contained in a small interval around $\bar{\gamma}$ as 
expected, since the two {\it high} sites are then almost unaffected by each other, each
essentially behaving like a one-site breather (3), (4). \\
Interestingly, a 2-site breather with the two {\it high} sites {\it in phase}
 with each
other can also be constructed in the model, but turns out to be always 
unstable. \\
The details of the calculations together with numerical and other results
on breather stability will be presented elsewhere.
\vskip 2cm 
\begin{figure}
\begin{center}
\epsfig{figure=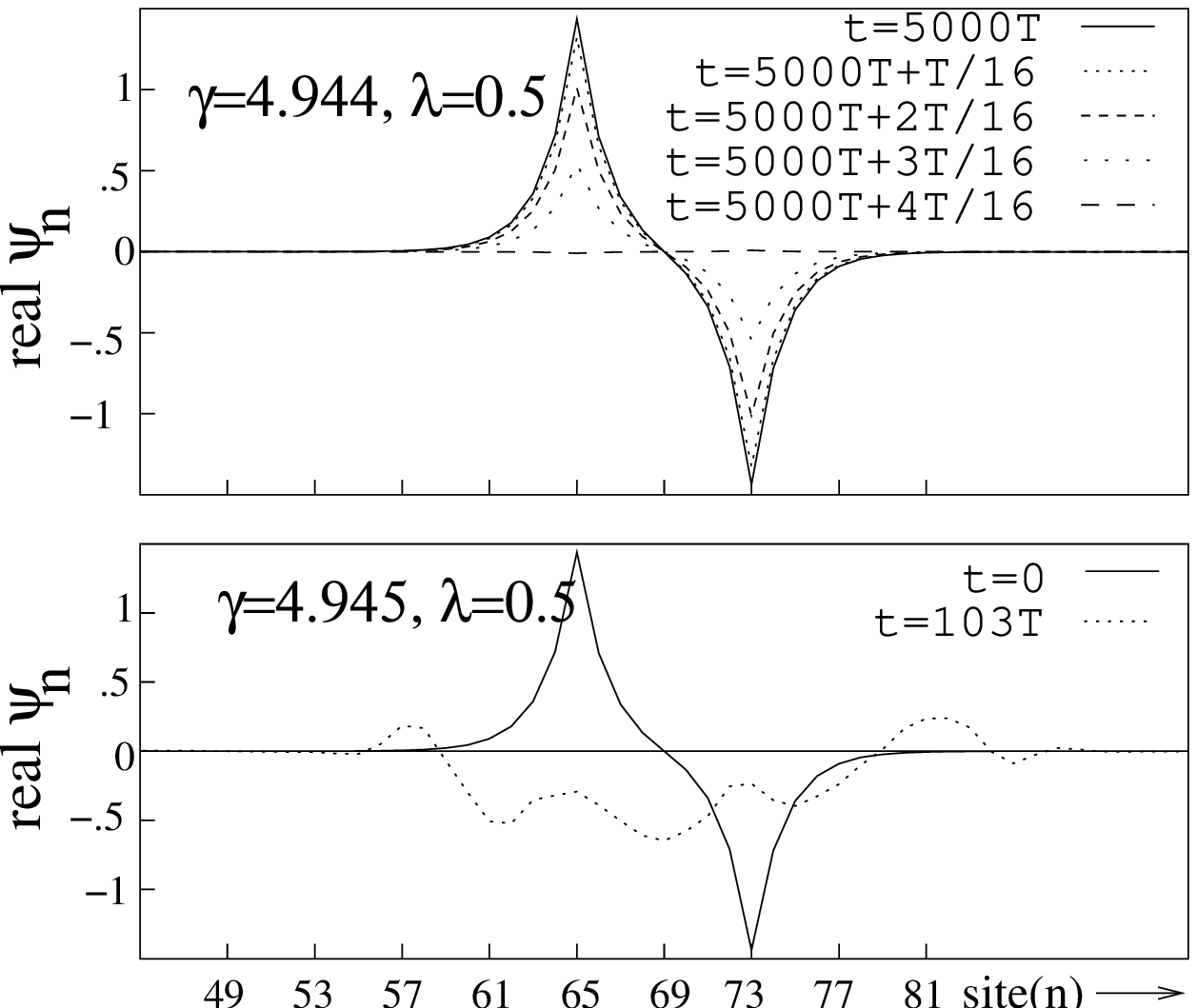,height=3cm,width=6cm}
\end{center}
\caption{Numerical integration following (1), (\ref{eq:two}) of an 
initial two-site breather profile given by (\ref{eq:eleven}): (a) $\gamma<\gamma_H
(\lambda)$, (b) $\gamma>\gamma_H
(\lambda)$.}
\end{figure}

\end{multicols}
\end{document}